\documentclass[12pt,preprint]{aastex}

\def\msun{{\rm\,M_\odot}}

\newcommand{\etal}{et al.\ }

\def\h2{${\rm\,H_2}$}

\makeatletter 

\def\etal   {{et~al.}\ }

\def\hkpc{{~h_{73}^{-1}\rm\,kpc}}
\def\hMpc{{~h_{73}^{-1}\rm\,Mpc}}

\def\msun{{\rm\,M_\odot}}

\def\vol#1  {{{#1}{\rm,}\ }}

\def\etal{et al.\ }

\newcount\refno
\newcount\rfno
\def\eq{$^{\the\refno\ }$\advance\refno by 1}
\def\ad{\advance\rfno by 1}

\def\clock{\count0=\time \divide\count0 by 60
     \count1=\count0 \multiply\count1 by -60 \advance\count1 by \time
     \number\count0:\ifnum\count1<10{0\number\count1}\else\number\count1\fi}

\def\myputfigure#1#2#3#4#5%
{\hskip0.03\textwidth\vskip#5pt
\makebox[0pt]{\hskip#2in
\includegraphics[width=#3\textwidth]{#1}}\vskip#4pt\hfill}

\begin{document}
\title{Galaxy Size Problem at \lowercase{$z=3$}: Simulated Galaxies Are Too Small}

\author{M.~Ryan~Joung\altaffilmark{1}, Renyue~Cen\altaffilmark{1}, \& Greg~L.~Bryan\altaffilmark{2}}  

\begin{abstract}
Using state-of-the-art adaptive mesh refinement cosmological hydrodynamic
simulations with a spatial resolution of proper $0.21\hkpc$ 
in refined subregions embedded within a comoving cosmological 
volume ($27.4\hMpc$)$^3$,
we investigate the sizes of galaxies at $z=3$ in the standard cold dark matter model
where reionization is assumed to complete at $z_{ri}\sim 6$.
Our simulated galaxies 
are found to be significantly smaller than the observed ones:
while more than one half of the galaxies 
observed by HST and VLT ranging from rest-frame UV to optical bands
with stellar masses larger than $2\times 10^{10}\msun$ have  
half-light radii larger than $\sim$$2\hkpc$, 
none of the simulated massive galaxies in the same mass range have 
half-light radii larger than $\sim$$2\hkpc$,
after taking into account dust extinction.
Corroborative evidence is provided by the rotation curves of the simulated galaxies 
with total masses of $10^{11}$-$10^{12}\msun$,
which display values ($300$-$1000~$km s$^{-1}$) 
at small radii ($\sim$$0.5\hkpc$) due to high stellar 
concentration in the central regions, 
larger than those of any well observed galaxies. 
Possible physical mechanisms 
to resolve this serious problem 
include: 
(1) an early reionization at $z_{ri}\gg 6$
to suppress gas condensation hence star formation,
(2) a strong, internal energetic feedback from stars or central black holes
to reduce the overall star formation efficiency,
or (3) a substantial small-scale
cutoff in the matter power spectrum.

\end{abstract}


\altaffiltext{1}{Department of Astrophysical Sciences, Princeton
  University, Peyton Hall, Ivy Lane, Princeton, NJ 08544}

\altaffiltext{2}{Department of Astronomy, Columbia 
  University, Pupin Physics Laboratory, New York, NY 10027}

\section{INTRODUCTION}

The standard cosmological model
has been remarkably successful in accounting for
observations on scales larger than galaxy sizes 
(Krauss \& Turner 1995; Ostriker \& Steinhardt 1995;
Bahcall \etal 1999;
Tegmark \etal 2004; Spergel \etal 2007).
We intend to test this same model  
with regard to galaxy formation and evolution,
a regime where astrophysical processes are important and
hence a detailed testing of the cosmological model becomes 
more intricate.
In this paper, the first of a series,
we focus on the sizes of galaxies at redshift $z=3$, 
including ``Lyman Break Galaxies'' (LBGs) (Steidel \etal 2003).
Previous works on this subject include those based on semi-analytic methods
(e.g., Mo \etal 1999; Somerville \etal 2001; Somerville \etal 2008);  
the observed size-mass and size-luminosity relations at $z=3$ were reproduced 
in these studies.

Here, we take a brute-force approach
using high-resolution adaptive mesh refinement (AMR)
cosmological simulations to minimize
the number of adjustable astrophysical parameters and thereby maximize
the predictability of the standard model.
A physical resolution of our simulations of $0.21\hkpc$ proper
at $z=3$ in refined subregions 
embedded within a comoving cosmological
volume ($27.4\hMpc$) permits, 
an accurate characterization of the sizes of galaxies at $z=3$.
More specifically, we compare the half-light radii of simulated and 
observed galaxies in terms of size-mass and size-luminosity relations.  
This comparison is motivated by a series of observations of $z \sim 3$ 
galaxies in the rest-frame UV (Giavalisco \etal 1996; Lowenthal \etal 1997; 
Ferguson \etal 2004; Giavalisco \etal 2008) 
and in the rest-frame optical (Trujillo \etal 2006; 
Toft \etal 2007; Zirm \etal 2007; Buitrago \etal 2008).

We find the simulated galaxies to be generally smaller than the 
observed galaxies at $z=3$ in the stellar mass range $\ge 10^{10.5}\msun$
where comparisons can be made.  
This problem may be related to the disk size problem 
(also called the ``angular momentum problem'') 
at $z=0$ (e.g, Navarro \etal 1995;
Navarro \& Steinmetz 1997;
Governato \etal 2004).
The galaxy size problem at $z=3$ found in the present work may also
be related to an apparent large excess 
of predicted but unobserved dwarf halos (Klypin \etal 1999; Moore \etal 1999)
and an over-concentration of dark matter in simulated 
dwarf galaxies on the scale of $\sim$$1\hkpc$ 
(Moore 1994; Flores \& Primack 1994; Burkert 1995;
McGaugh \& de Blok 1998; 
Kravtsov \etal 1998;
Moore \etal 1999).
Physical processes at high redshift that may be responsible
for the eventual resolution of this problem
should be manifested more clearly
at $z=3$, as they are unaffected by additional complications that 
may occur at lower redshifts.
Therefore, one may be able to obtain important ``cleaner" clues
to the nature of the dark matter and/or important astrophysical 
processes at high redshifts 
by studying 
$z=3$ galaxies.
Moreover, combining observations at both $z=3$ and $z=0$
may provide still more powerful constraints.

\section{SIMULATION AND ANALYSIS METHODS}

We perform cosmological simulations with the adaptive mesh refinement (AMR) 
Eulerian hydro code, Enzo (Bryan 1999; Norman \& Bryan 1999; O'Shea et al. 2004).  
First, we ran a low resolution simulation with a periodic box of $27.4\hMpc$ comoving on a side 
in a $\Lambda$CDM universe with cosmological parameters consistent with the WMAP3 results: 
($\Omega_m$, $\Omega_{\Lambda}$, $\Omega_b$, $h$, $\sigma_8$, $n_s$) = (0.24, 0.76, 0.042, 0.73, 0.74, 0.95).  
We identified virialized dark matter halos in this simulation at $z=3$ 
and resimulated 11 of the most massive 20 halos 
in a suite of five high resolution simulations embedded within the same ($27.4\hMpc$)$^3$ comoving box 
using the multimass initialization technique.
The five high resolution subregions have comoving volumes ranging
from $\sim$($3.4\hMpc$)$^3$ to $\sim$($8.8\hMpc$)$^3$. 
Within the high-resolution regions, the cell size of the root grid is $53.5\hkpc$ 
and additional grid refinements are allowed to reach 
a maximum level of $l_{max}=6$, resulting in the maximum spatial resolution 
of $0.84\hkpc$ (comoving) or $0.21\hkpc$ (proper) at $z=3$,  
while the rest of the box is evolved at a lower resolution of $214\hkpc$. 
The dark matter particle mass in the high-resolution region is $4.6 \times 10^6$ M$_{\odot}$.
The simulations include a metagalactic UV background (Haardt \& Madau 1996), 
a diffuse form of photoelectric and photoionization heating 
(Abbott 1982; Joung \& Mac Low 2006)
and shielding of UV radiation by neutral hydrogen (Cen \etal 2005). 
They also include cooling due to molecular hydrogen (Abel et al. 1997)
and  metallicity-dependent radiative cooling 
(Cen \etal 1995) extended down to $10~$K (Dalgarno \& McCray 1972). 
Star particles are created in cells that satisfy a set of criteria for 
star formation proposed by Cen \& Ostriker (1992).  
A stellar particle of mass
$m_{*}=c_{*} m_{\rm gas} \Delta t/t_{*}$ is created
(the same amount is removed from the gas mass in the cell),
if the gas in a cell at any time meets
the following three conditions simultaneously:
(1) contracting flow, (2) cooling time less than dynamic time, and  (3)
Jeans unstable,
where $\Delta t$ is the time step, $t_{*}={\rm max}(t_{\rm dyn}, 10^5$yrs),
$t_{dyn}=\sqrt{3\pi/(32G\rho_{tot})}$ is the dynamical time of the cell,
$m_{\rm gas}$ is the baryonic gas mass in the cell and
$c_*=0.03$ is star formation efficiency.
Each star particle is tagged with its initial mass, creation time, and metallicity; 
star particles typically have masses of $\sim$$10^5\msun$.
Star formation and supernovae feedback are modeled following Cen \etal (2005)
with $e_{SN}=3\times 10^{-6}$.
Feedback energy and ejected metals are distributed into 
27 local gas cells centered at the star particle in question, weighted by the specific volume of each cell.
The temporal release of metal-enriched gas and thermal energy at time $t$
has the following form: 
$f(t,t_i,t_{dyn}) \equiv (1/ t_{dyn})
[(t-t_i)/t_{dyn}]\exp[-(t-t_i)/t_{dyn}]$,
where $t_i$ is the formation time of a given star particle.
The metal enrichment inside galaxies and in the intergalatic medium 
is followed self-consistently in a spatially resolved fashion (Cen \etal 2005).

We identify virialized objects in our high resolution simulations using the HOP algorithm (Eisenstein \& Hut 1998),
which is tested to be robust.
We find 49 halos with virial masses $>5 \times 10^{10}$ M$_{\odot}$ 
to compare with observations.

The light distribution is computed from the star particles using the GISSEL stellar
synthesis code (Bruzual \& Charlot 2003).  
We calculate the luminosities of the simulated galaxies in various 
rest-frame UV and optical bands where observations are available for 
$z \sim 3$ galxies with measured half-light radii (see Fig. 2).  To 
obtain half-light radii in the right wavelength range, we placed the same 
filter (blueshited to z=3) as used in each sample of observed galaxies;  
this was necessary because the sizes of simulated galaxies vary depending 
on the observed band, as shown in the size-luminosity plot.  
We followed the procedure described in Appendix B of Rudnick et al. (2003) 
to compute luminosities of the model galaxies in each given 
band.\footnote{For photon counting devices, the filter transmission curves 
$T(\lambda)$ in Rudnick et al. (2003) should be replaced by $T(\lambda) \lambda$.}  The stellar 
mass of each simulated galaxy is equal to the sum of the masses of the 
star particles located 
within 15\% of the virial radius of the galaxy at z=3.  For the observed 
galaxies, we adopt the stellar masses reported in the papers referenced 
in Figure 2 and, where appropriate, convert apparent magnitudes to 
luminosities in the wavelength range of the given filter blueshifted to 
z=3.
We adopt an approximate model for dust extinction following Binney \& Merrifield (1998) but assume that dust attenuation is proportional to the metal column density rather than the total hydrogen density and correct for depletion of refractory elements (Zn) onto dust 
grains parametrized by $f_{Fe}$, fraction of iron in dust (Vladilo \& Peroux 2005):
$A_V = \Sigma_Z~f_{Fe}/(4\times 10^{19}~m_p F$ cm$^{-2})$ mag, 
where $m_p$ is the proton mass, $\Sigma_Z$ is the mass column 
density of metals in front of a given star particle. 
The factor, $F=3.7$, is chosen so that $E(B-V) = 0.3$ (Pettini \etal 1998; 
Shapley \etal 2003; Steidel \etal 2003), corresponding to $A_{1500\AA} 
\approx 1.5$ and an escape fraction in the rest-frame $1500\AA$ is $\sim$26\%. 
We choose a relatively high value of $E(B-V)$ to strengthen our conclusion 
that, even though dust extinction acts to increase half-light radii of the 
galaxies, our simulated galaxies are still generally smaller than the 
observed ones.  
The conversions from $A_V$ to $A$ at other bands are based on 
the dust extinction law proposed by Calzetti \etal (2000). 
All half-light radii are measured directly using projected 2D maps.

\section{RESULTS}

Figure 1 shows the projected 
stellar mass density of a region of comoving size
($1.4\hMpc$)$^2$ with a depth of comoving $3.3\hMpc$ at $z=3$,
cut out from our largest high resolution simulation sub-volume of size $\sim$($8.8\hMpc$)$^3$.
The insets show magnified images of the
four most massive halos in the displayed region 
in their projected luminosity density distributions. 
For each galaxy, the
images show, from left to right, intrinsic (before dust attenuation is
applied), apparent (after dust attenuation), and smoothed (after dust
attenuation and PSF filtering) luminosity densities, respectively, all in
the observed $I$-band.\footnote{We use the response curve for the F814W filter on HST's WFPC2.}
On the one hand, we see that our high resolution permits formation of extremely dense
structures.
On the other hand, many rich and salient features produced by cosmological processes,
such as mergers and tidal tails, are clearly visible and striking.
Comparing the leftmost and middle pictures of each row of the insets 
suggests that dust extinction significantly affects 
the observed luminosity density distribution 
in the observed $I$-band. 
This can be understood simply as a consequence of the empirical Schmidt-Kennicutt relation (Kennicutt 1998): 
gas surface density increases monotonically 
as star formation rate density increases; and we assumed that 
dust extinction is proportional to the metal surface density 
hence gas surface density (for a constant metallicity).

Figure 2 shows the size-mass (top) and size-luminosity relations (bottom)
of the simulated and observed galaxies.  The ``intrinsic" half-light radii of the 
simulated galaxies are displayed as ``x", while dust extinction-applied 
half-light radii are shown as filled circles where the shade in grey scale 
indicates the amount of dust extinction in the rest-frame $V$-band (darker 
for higher extinction; $0.35 \lesssim A_V \lesssim 2.5$).  Each sample of 
observed galaxies is shown by distinct symbols as indicated.
The observed $z \sim 3$ galaxies include the rest-frame UV samples from 
Lowenthal \etal (1997) and Giavalisco \etal (2008) as well as rest-frame 
optical samples from Trujillo \etal (2006), Toft \etal (2007), Zirm 
\etal (2007) and Buitrago \etal (2008).
In all panels, the intrinsic sizes of the simulated galaxies lie mostly between 0.2 and 
$0.5 \hkpc$, while the dust extinction-applied sizes can be significantly larger.  
This increase in apparent sizes is, on average, 
greater at shorter wavelengths and for more massive and luminous galaxies, 
and may be partly responsible for the observed relation between star formation 
and size at $z \sim 2.5$ (Toft et al. 2007; Zirm et al. 2007).  
Remarkably, this trend is also seen in the sizes of the simulated galaxies 
measured in the rest-frame optical bands. 

In the top display of Figure 2, 
the observational data are only available for galaxies with 
stellar masses of $\ge 2\times 10^{10}\msun$.
In this mass range, there are more than one half of the observed galaxies
having the half-light radii (all measured in the rest-frame optical) 
greater than $2 \hkpc$,
whereas all of the simulated galaxies in the same mass range
have the half-light radii smaller than $2 \hkpc$.
A significant fraction of these relatively massive galaxies 
that are actively star-forming (SFR $> 10$ M$_{\odot}$ yr$^{-1}$) 
may be identified with the LBGs (Steidel \etal 1996).

This discrepancy in half-light radii between simulated 
and observed galaxies is also seen in the size-luminosity relation
at the bottom display of Figure 2, where luminosities at several different 
observed bands are shown in four separate panels.
Here, the half-light radii of the simulated 
galaxies again occupy the low-end of the distributions compared to 
galaxies observed in the HST WFPC2 F814W filter 
(Lowenthal \etal 1997),
in the VLT ISAAC K$_s$ band (Trujillo \etal 2006;
Toft \etal 2007).
However, the situation here is slightly more complicated 
in that we find apparent agreement between simulated galaxies
and observations of Giavalisco \etal (2008) 
in the HST ACS F850LP filter 
and, to some degree, of Buitrago \etal (2008) in the VLT ISAAC K$_s$ band.
Whether the 
discrepancy among observed sizes is real or an artifact of the different 
source extraction and fitting algorithms employed is an open 
question.\footnote{We note, however, that the measured half-light radii 
in Giavalisco et al. (2008) 
-- based on the SExtractor program -- may be biased low compared to size 
measurements derived from fits to Sersic profiles, for the faint galaxies in 
their sample (M. Giavalisco, private comm.).}

Overall, we conclude that there is fairly strong indication
that the simulated massive galaxies are too small 
compared to their observed counterparts at $z \sim 3$,
from rest-frame UV to rest-frame optical bands.
This suggests that the concurrent star formation activities and the overall 
stellar mass distributions in the simulated 
galaxies at $z=3$ are both too concentrated near the galactic centers.
Concentration of stellar mass may 
be shown in an alternative way using rotation curves.
Figure 3 shows rotation velocity curves for the three top galaxies 
of total mass of $10^{11}-10^{12}\msun$.
These curves seem to peak at too high a value ($300$-$1000~$km s$^{-1}$)
at small radii ($\sim$$0.5\hkpc$).

It is prudent to ask if our results depend on limited resolution.
To address this point, we ran higher resolution simulations (with the 
smallest cell size of  0.11 proper$\hkpc$ or twice the linear 
resolution as in our fiducial models) of 2 of the 5 subvolumes,
or 8 of the 49 galaxies discussed in this Letter.  In all of these cases,
we find that both the intrinsic half-mass and intrinsic half-light radii
are significantly smaller (by nearly a factor of 2), indicating that our
results have not yet converged in terms of obtaining the absolute sizes
of the high-z galaxies.
We find that the dust attenuated half-light
radii with the higher resolution simulations are slightly smaller (10-30\%)
than those from lower resolution simulation. 
While the absolute sizes have not converged,
these tests support and strengthen our
conclusion that the simulated galaxies at $z=3$ are too small
compared to their observed counterparts.

\section{DISCUSSION AND CONCLUSIONS}

The standard cold dark matter cosmological model is in good agreement 
with a rich set of observations on large scales.
Our main purpose is to systematically examine 
it, through a series of papers,
in the context of galaxy formation and evolution,
a regime where, relatively speaking, 
it has not been seriously contested.
Because astrophysical processes tend to play a progressively 
more important role at smaller scales,
in particular, on galactic scales and smaller,
it is vital to employ as few 
adjustable astrophysical parameters as possible,
to have a true test of the cosmological model.

In this paper we focus on the sizes of galaxies, including
LBGs (Steidel \etal 2003) at redshift $z=3$,
using state-of-the-art AMR cosmological hydrodynamic
simulations with a spatial resolution of proper $0.21\hkpc$ 
in five refined subregions embedded within a comoving cosmological volume ($27.4 \hMpc$)$^3$.
We find that, taking into account dust extinction, 
the computed distribution of half-light radii of simulated galaxies 
in the rest-frame $I$- and $V$-bands occupy only the low-end of the 
observed size distributions.  
While none of the simulated massive galaxies have 
half-light radii larger than $\sim$$2\hkpc$, 
more than one half of the observed galaxies have  
half-light radii exceeding that value. 
We note that the intrinsic 
(i.e., in the absence of dust extinction and finite instrumental resolution)
half-light (and half-stellar-mass) radii are smaller at $\sim 0.3\hkpc$
and our resolution tests indicate that they become still smaller
with higher resolution, hence strengthening our conclusions. 
Consistent with this apparent discrepancy between simulated and observed
LBGs, the rotation curves of the simulated galaxies
of total masses of $10^{11}$-$10^{12}\msun$
have unusually high values ($300$-$1000~$km s$^{-1}$) 
at small radii ($\sim$$0.5\hkpc$).
Such rotation curves are not seen in local galaxies for which accurate 
measurements are available,
although no direct comparison with 
observed high-redshift galaxies 
can be properly made at this time.

These discrepancies appear to originate from 
stellar masses that are too highly concentrated in small ($< 1\hkpc$) central regions
of the simulated galaxies.
This may be caused by an over-abundance of smaller galaxies
that formed at high redshifts and subsequently sank to the centers
via dynamical friction {\it or} by vigorous {\it in situ} star formation in the central regions.
Likely, 
any potential viable solution to this apparent problem would have to reduce the amount 
of stars that formed.
Possible physical mechanisms include: 
(1) an early reionization with $z_{ri}\gg 6$
to suppress gas condensation that will reduce earlier star formation
(e.g., Bullock, Kravtsov, \& Weinberg 2000),
(2) a strong, internal energetic feedback from stars or central black holes
to reduce the overall star formation efficiency (e.g., Sommer-Larson \etal 2003;
Governato \etal 2007),
or 
(3) a substantial small-scale
cutoff in the matter power spectrum,
for example, if the dark matter particles are warm rather than cold
(e.g., Hogan \& Dalcanton 2000; Sommer-Larsen \& Dolgov 2001;
Bode, Ostriker \& Turok 2001)

Since the age of the universe at $z=3$ is only about $1/6$ of the present age, 
a successful resolution to the galaxy size problem at $z\sim 3$
may provide important ``cleaner" clues
to the nature of the dark matter and/or important astrophysical 
processes at high redshifts.
Moreover, combining observations at both $z=3$ and $z=0$
may provide still more powerful constraints.

\acknowledgements{
We thank A. Burkert, Y.-T. Lin, M.-M. Mac Low, 
L. Mayer, B. Moore, R. Overzier, A. Shapley, and P. van Dokkum 
for helpful discussions and comments; 
F. Buitrago, M. Giavalisco, M. Kriek, G. Rudnick, and I. Trujillo 
for generous help with observational data; and 
M.-S. Shin and K. Nagamine 
for help with the galaxy spectrum code.
We are grateful to M. Giavalisco and his colleagues for
making their new GOODS-S sample available to us prior to publication.
We are indebted to an anonymous referee for many useful suggestions
that help improve the paper.
We thank M. Norman and his team 
in University of California at San Diego
for useful assistance in code development.
We gratefully acknowledge financial support by
grants AST-0507521, NNG05GK10G and NNX08AH31G.
The simulations were performed at NCSA with computing time provided by
LRAC allocation TG-MCA04N012.
}

\clearpage

\begin{figure*}
\begin{center}
\resizebox{7.5in}{!}{\includegraphics[angle=0]{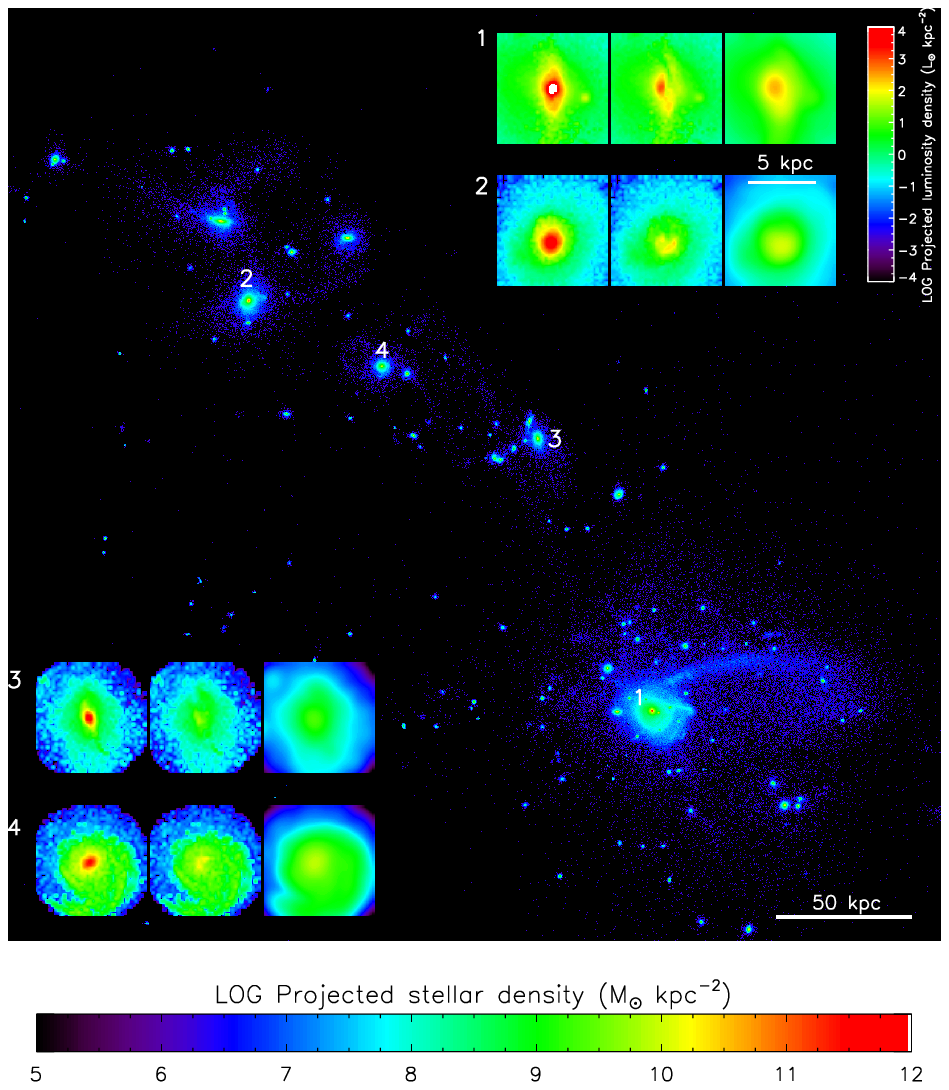}} 
\end{center}
\caption{
The large image displays the projected 
stellar mass density of a region of comoving size
$(1.4\hMpc$)$^2$ with a depth of comoving $3.3\hMpc$ at $z=3$,
cut out from one of our simulation volumes.
The insets zoom in on the
four most massive galaxies in volume region
in terms of virial mass and show
their projected luminosity density distributions.  For each galaxy, the
images show, from left to right, intrinsic (before dust attenuation is
applied), apparent (after dust attenuation), and smoothed (after dust
attenuation and PSF filtering) luminosity densities, respectively, all in
the observed $I$-band.  
A Gaussian filter with FWHM $=0''.125$ is applied for the PSF.
The bars indicate lengths in proper$\hkpc$.
}
\label{fig:1}
\vskip7pt
\end{figure*}

\begin{figure*}
\begin{center}
\resizebox{3.7in}{!}{\includegraphics[angle=0]{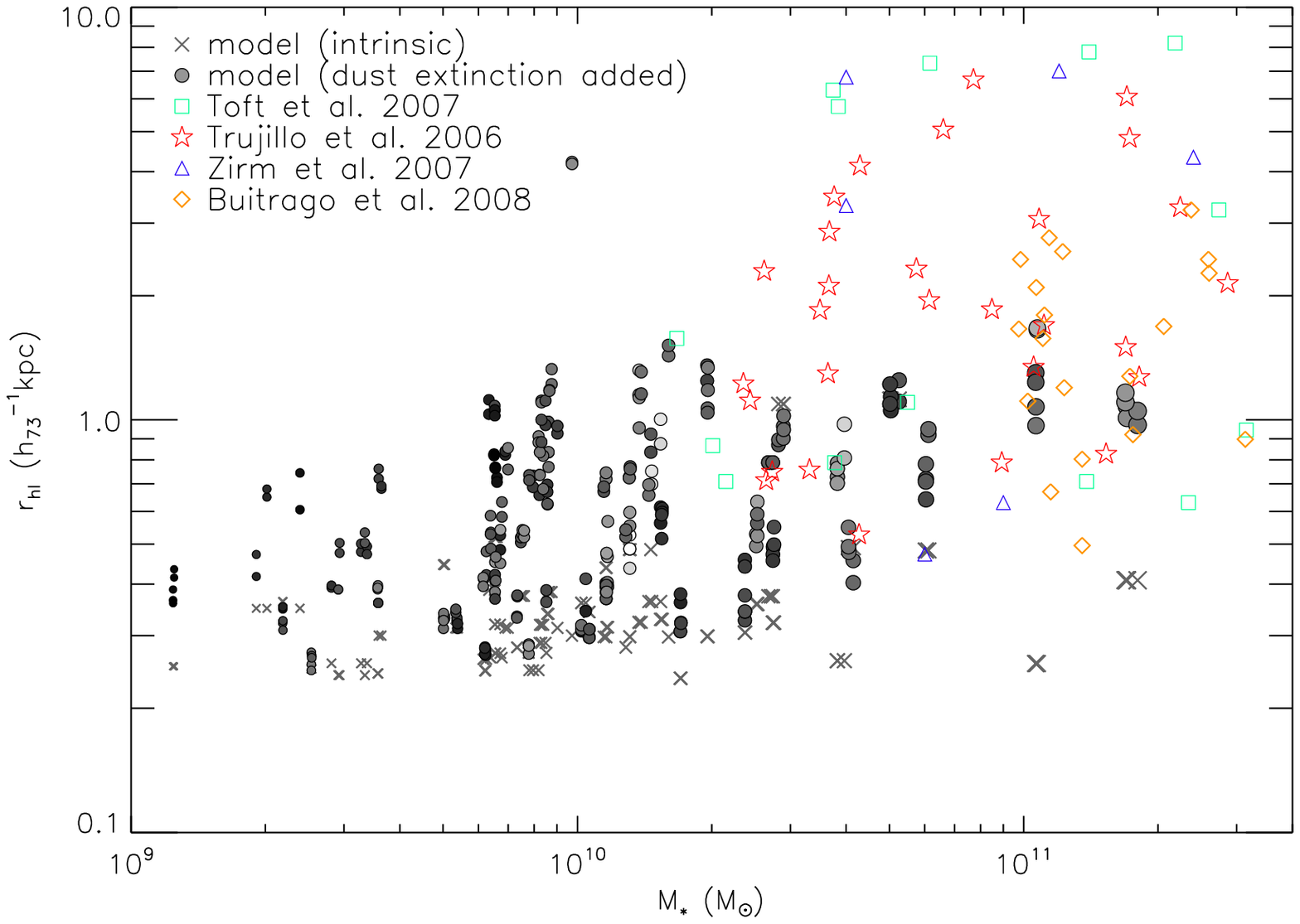}}
\resizebox{4.0in}{!}{\includegraphics[angle=0]{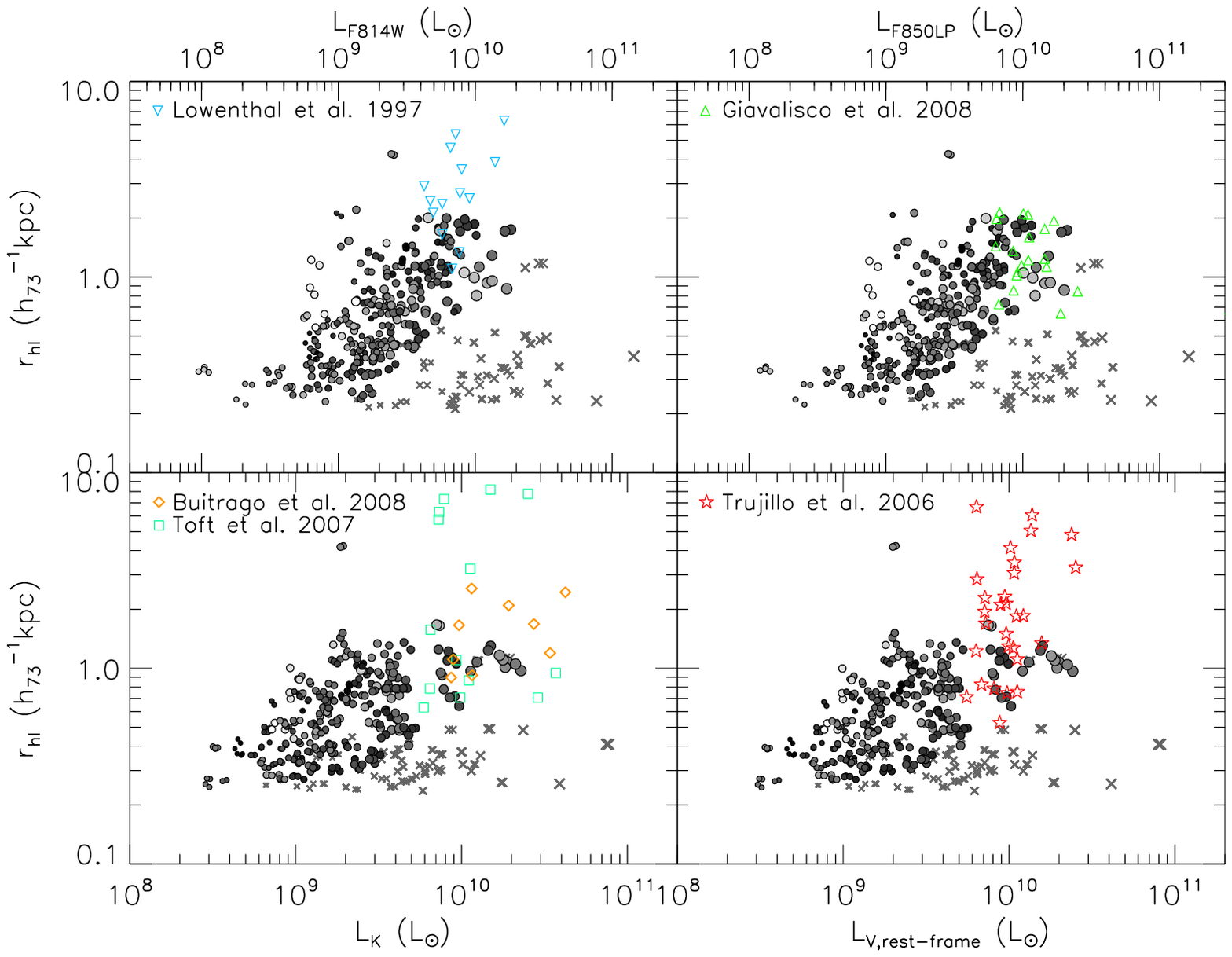}}
\end{center}
\caption{
Size-mass ({\it top}) and size-luminosity ({\it bottom}) relations 
of the simulated and observed galaxies.  The ``intrinsic" half-light radii 
of the simulate galaxies (along six orthogonal projection directions) are 
displayed as crosses, while dust extinction-applied 
half-light radii are shown as filled circles where the shade in grey scale 
indicates the amount of dust extinction in the rest-frame $V$-band (darker 
for higher extinction; $0.35 \lesssim A_V \lesssim 2.5$).  
In the top display the observed 
galaxies that appear in the size-mass relation are based on measurements 
in the rest-frame optical.  
In the bottom display 
the size-luminosity relations are shown for 
both rest-frame UV ({\it top panels}) and rest-frame optical ({\it 
bottom panels}). 
The observed $z \sim 3$ galaxies include the rest-frame UV samples from 
Lowenthal \etal (1997) using HST WFPC2 F814W 
filter ($\lambda_c=8203\AA$ and FWHM=$1758\AA$) 
and from Giavalisco \etal (2008) using HST ACS F850LP 
filter ($\lambda_c=8950\AA$ and FWHM=$900\AA$),
rest-frame optical samples from Trujillo \etal (2006), 
Toft \etal (2007) and Buitrago \etal (2008)
using VLT ISAAC K$_s$ band ($\lambda_c=21600\AA$ and FWHM=$2700\AA$)
and from Zirm \etal (2007) using HST WFPC2 F160W 
filter ($\lambda_c=16089\AA$ and FWHM=$4010\AA$).
}
\label{fig:2}
\vskip7pt
\end{figure*}

\begin{figure*}
\begin{center}
\resizebox{6.0in}{!}{\includegraphics[angle=0]{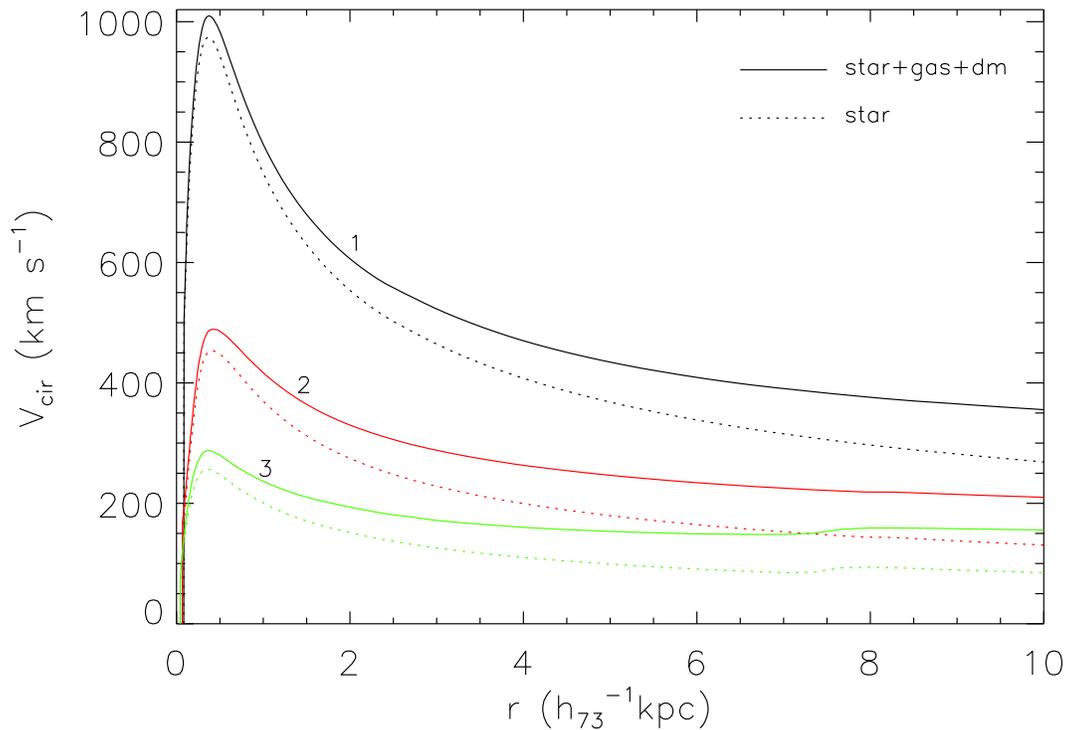}}
\end{center}
\caption{
Rotation velocity curves for the three galaxies labeled ``$1$'', ``$2$''
and ``$3$'' in Fig. 1.  The solid curves represent rotation velocities
due to all the matter within a given galactocentric radius,
while the dotted curves show those due to stellar mass only.  
The virial masses of the three galaxies
are 
$8\times 10^{11}$, 
$4\times 10^{11}$
and 
$1\times 10^{11}\msun$, respectively.
}
\label{fig:3}
\vskip7pt
\end{figure*}

\end{document}